\documentclass{amsart}

\usepackage{amsmath,amssymb,amscd,graphicx,epsfig}
\everymath{\displaystyle}

\newtheorem{teo}{Theorem}[section]
\newtheorem{lem}[teo]{Lemma}

\newtheorem{prop}[teo]{Proposition}
\newtheorem{lem-defi}[teo]{Lemma-Definition}

\newcommand{\mr}{\Bbb{R}}

\newcommand{\mg}{\Bbb{G}}

\newcommand{\mz}{\Bbb{Z}}

\newcommand{\B}{{\mathcal B}}
\newcommand{\Cc}{{\mathcal C}}
\newcommand{\D}{{\mathcal D}}
\newcommand{\Ee}{{\mathcal E}}
\newcommand{\Ff}{{\mathcal F}}
\newcommand{\G}{{\mathcal G}}
\newcommand{\Hh}{{\mathcal H}}

\newcommand{\Ll}{{\mathcal L}}

\newcommand{\Ss}{{\mathcal Q\mathcal C}}
\newcommand{\Tt}{{\mathcal T}}

\begin{document}

\title  { Minimal configurations\\ for  the Frenkel-Kontorova model \\on  a  quasicrystal}

\pagestyle{headings}
\noindent \author{  by\\Jean-Marc Gambaudo,\, Pierre Guiraud and Samuel Petite\\ \,\\ \,\\ \,\\ \, \\ \,\\ \, \\ \,\\ \,  }

\begin{abstract} In this paper, we consider  the Frenkel-Kontorova model of a one dimensional  chain of atoms submitted to a potential. This potential splits into an interaction potential and a potential induced by an underlying substrate which is a quasicrystal. Under standard hypotheses, we show that every minimal configuration has a rotation number, that the rotation number varies continuously with the minimal configuration, and that every non negative real number is the rotation number of a minimal configuration. This generalizes well known results obtained by S. Aubry and P.Y. le Daeron   in the case of a crystalline substrate. 
\end{abstract}
\noindent \address{ {\it J.-M. Gambaudo: } Centro de Modelamiento Matem\'atico, U.M.I. CNRS 2807, Universidad de Chile, Av. Blanco Encalada 2120, Santiago, Chile.}
\email{gambaudo@dim.uchile.cl}

\noindent \address{{\it P. Guiraud:}  Departamento de Ingenier\'{\i}a Matem\'atica,
Fac. Ciencias F\'{\i}sicas y Matem\'a\-ticas, Universidad de Chile,
Av. Blanco Encalada 2120 5to piso, Santiago, Chile.}
\email{pguiraud@dim.uchile.cl}

\address{{\it S.  Petite:}  Institut
de Math\'ematiques de Bourgogne, U.M.R. CNRS 5584, Universit\'e de
Bourgogne, U.F.R. des Sciences et T\'echniques, B.P. 47870-\,\,
21078 Dijon Cedex, France. }
\email{samuel.petite@u-bourgogne.fr}
\date{\today}

\maketitle
\markboth{ J.-M. Gambaudo, P. Guiraud  and S. Petite}{The Frenkel-Kontorova model on a quasicrystal}
\eject
\section{Introduction}\label{Intro}

The Frenkel-Kontorova  model   \cite{FK}  describes the physical situation of a layer of a material over a substrate of other material (see for instance \cite{book}). In the one dimensional case, the  layer of material is described  by the configurations of a bi-infinite chain of particles on the real line. These configurations are parametrized by a  bi-infinite non decreasing sequence ${(\theta_n)}_{n\in\mz}$ of real  numbers, where 
$\theta_n$  represents the position of the particle labeled by $n$.  

\noindent The potential energy of the chain  reads:
$$\Ee({(\theta_n)}_{n\in\mz})\, =\, \sum_{n\in \mz}\,U (\theta_n - \theta_{n+1})\, +\, V(\theta_n),$$
where  $U$ describes   the interaction between particles (only interactions with the nearest neighbors are considered),  and $V$ is  a potential induced by the substrate and  depends on its nature. 

\noindent The following standard extra asumptions are made on $U$ and $V$:
\begin{itemize}
\item {\it Smoothness:} the functions $U$ and $V: \mr\to \mr$ are $C^2$;
\item {\it Convexity:}
$U''(x) >0,\quad \forall\, x\in \mr;$
\item {{\it Behavior at $\infty$}}:
  $\lim_{x\to \pm\infty} \frac{{U(x)}}{\vert x\vert}\,= \,+\infty$.
\end{itemize}

Even if the above sum is only formal, 
 it is possible to look for  equilibrium configurations which minimize locally the energy (ground states). More precisely
 let us consider the function $\Hh:\mr \times\mr\to \mr$ defined by:
$$\Hh(\theta, \theta')\, =\,U(\theta - \theta' ) \, +\, V(\theta).$$ 
For a configuration $ {(\theta_n)}_{n\in \mz}$, let us set:
$$\Hh_p(\theta_i, \theta_{i+1}, \dots, \theta_{i+p}) \, =\, \sum_{j=0}^{j =p-1} \Hh(\theta_{i+j}, \theta_{i+j+1}).$$
We say that the {\it segment}    $  (\theta_i, \theta_{i+1}, \dots, \theta_{i+p} )$  of the configuration ${(\theta_n)}_{n\in \mz}$ is {\it minimal}
if 
$$\Hh_p(\theta_i, \theta_{i+1}, \dots, \theta_{i+p}) \, \leq \, \Hh_p(\theta'_i, \theta'_{i+1}, \dots, \theta'_{i+p}),$$ for any other segment $ (\theta'_i, \theta'_{i+1}, \dots, \theta'_{i+p})$ such that $\theta'_i = \theta_i$ and $ \theta'_{i+p} = \theta_{i+p}$. 
\noindent A configuration ${(\theta_n)}_{n\in\mz}$ is {\it minimal} if all its segments are minimal. 

The substrate is a {\it crystal} when the configuration of  the chain  of atoms it is made of, is an increasing sequence $\Ss = {(s_n)}_{n\in \mz}$  such that  there exists $q\in \mz^+$  and $L>0$ verifying:
$$s_{n+ q}\,=\, s_n \,+\,  L,\quad  \forall\  n\in \mz.$$ In this case it is natural to consider that   a potential $V$  {\it associated} with  the crystal  $\Ss$ is a periodic $C^2$-function with period $L$:
$$V(\theta + L)\,=\, V(\theta),\quad \forall\  \theta\in \mr.$$
This situation when the substrate potential is periodic has been described by S. Aubry and P. Y.  Le Dearon. Their seminal work \cite{AD},  together with the independent approach of J. Mather \cite{Mather}, gave rise to the so called {\it Aubry-Mather  theory}, which yields in particular a good understanding of minimal configurations. 

\noindent Let $\rho\in \mr$, a  configuration $ {(\theta_n)}_{n\in \mz}$ has a  {\it rotation number} equal to $\rho$ if the limit:
$$\lim_{n\to \pm\infty}\frac{ \theta_n}{n}\, =\, \rho.$$
Let us remark that the inverse of the rotation number  can be interpreted as a particle density. 

\noindent Aubry and le Daeron proved in particular that any minimal configuration has a rotation number,  that the rotation number is a continuous function when defined on the set of minimal configurations  equipped with the product topology,  and that any positive real number is the rotation number for some minimal configuration\footnote{Actually Aubry-Mather theory says much more about the combinatorics of minimal configurations when projected on a circle with length $L$.}.

\bigskip
 {\sl The aim of this paper is to consider  the case when the substrate is a quasicrystal in order to derive, in this more general setting, a similar description of the set of minimal configurations. }

\bigskip

To fix notations and definitions, let us consider a bi-infinite substrate chain of atoms represented by its  configuration $ {(s_n)}_{n\in \mz}$. 
Two segments $(s_n, \dots, s_{n+p})$ and  $(s_q, \dots, s_{q+p})$ are said {\it equivalent} if there exists $\tau \in \mr$ such that:
$$s_{q+i}\, =\,s_{n+i}\, +\, \tau, \quad\forall \, i\, =\, 0, \dots, p.$$
The chain $\Ss =  {(s_n)}_{n\in \mz}$ is a {\it quasicrystal} if the following properties are satisfied\footnote{See Proposition \ref{min} for a dynamical interpretation.}(see for instance \cite{LP}):
\begin{itemize}
\item {\it Finite local complexity} 

\noindent For any $M>0$, the chain possesses only finitely many equivalence classes of segments with diameters smaller than $M$. 
\item {\it Repetitivity} 

\noindent For any segment $S$ in the chain, there exists $R>0$ such that any ball  with radius $R$ contains a segment equivalent to $S$. 
\item {\it Uniform pattern  distribution} 

\noindent For any segment $S$ in  the chain,  and for any point $x \in \mr$, the quantity 
$$\frac{n(S, x, M)}{M} $$ converges when $M\to +\infty$ uniformly in $x$ to a limit $\nu(S)$ that does not depend on $x$,  where  $n(S, x, M)$ denotes the number of segments equivalent to $S$ in the interval $[x, x+M]$.  \end{itemize}

\noindent Notice that a crystal (with period  $L$) is a quasicrystal and in this particular case, for each segment $S$ in $\Ss$, one has:
$$\nu(S)\, =\, \frac{p(S)}{L},$$ where $p(S)$ stands for the number of segments equivalent to $S$ in a period $L$.

 For any $R>0$, a function $V_\Ss: \mr\to \mr$ is a {\it potential with range $R$} associated with a quasicrystal $\Ss$ if 
 for each pair of points $x$ and $y$ in $\mr$  such that 
 $$\Ss\cap B_R(x)\,-\, x =\, \Ss\cap B_R(y) \,  -\, y,$$  we have:
$$V_\Ss(x)\,  =\, V_\Ss(y), $$
where $B_M(z)$ stands for the ball with center $z$ and radius $M$. 
Whenever $\Ss$ is a crystal with period $L$,   it is clear that a potential  with range $R>0$ associated with this  crystal is a periodic potential with period $L$.   

\noindent We call  {\it short range potential}  associated with a quasicrystal $\Ss$ a potential with range $R,$ for some $R>0$. 

\begin{figure}\label{fibo}
\epsfxsize=8truecm
\centerline{\epsfbox{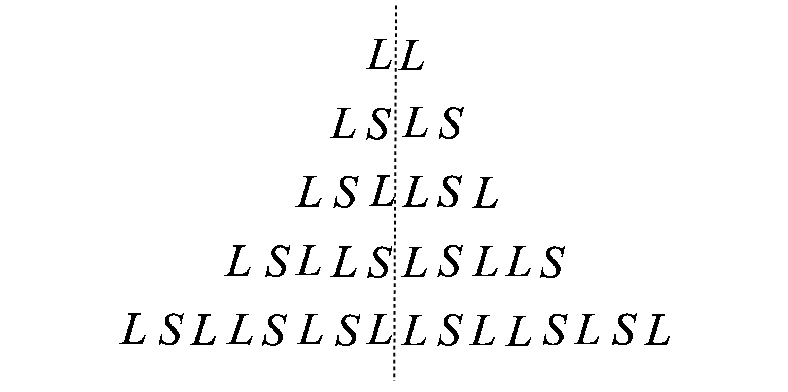}}
\caption{Construction of the Fibonacci chain}
\end{figure}

\noindent{\bf Example:}
A standard example of quasicrystal is given by the Fibonacci sequence. Consider the set  $\G$ of configurations ${(s_n)}_n$  such that:
\begin{itemize}
\item $s_0$ is located at $0$;
\item the lengths of the intervals $[s_n, s_{n+1}]$ have two possible sizes: either large and equal to $L$ or small  and equal to $S$.
\end{itemize}
The substitution:

$$
\quad \left\{
\begin{array}{l}L\, \to \, LS\\
S\, \to\, L\\
\end{array}
\right.
$$
induces a map $\Psi$  on $\G$ defined as follows:

\noindent  For a sequence ${(s_n)}_n$ in $\G$, consider the sequence of lengths ${(l_n)}_n \in \{L, S\}^\infty$  defined by 
$l_n =  s_{n+1} - s_n, \, \forall n\in \mz$. Applying to each $l_n$ the substitution rule we get a new sequence 
${(l'_n)}_n \in  \{L, S\}^\infty$. The new configuration ${(s'_n)}_n = \Psi({(s_n)}_n)$ is obtained by setting:
\begin{itemize}
\item  $s'_0 =0;$
\item$s'_{n+1} = s'_n + l'_n,\, \forall n\in \mz.$
\end{itemize}
Starting with the  equidistributed configuration ${(s^0_n)}_n$,  where $s_{n+1} - s_n = L, \forall n\in \mz, $ it is easy to check that the sequence of configuration ${(\Psi^k({(s^0_n)}_n))}_k$ converges when $k\to +\infty$ (for the product topology) to a configuration ${(s^\infty_n)}_n$ . This configuration is on the  one hand a quasicrystal and on the other hand a periodic point with period 2 of the operator $\Psi$.  This quasicrystal is called the {\it Fibonacci chain}  (See Figure \ref{fibo}).

\begin{figure}
\epsfxsize=8truecm
\centerline{\epsfbox{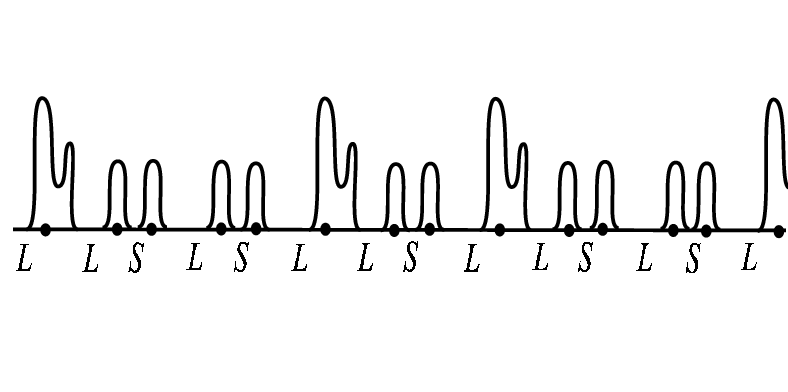}}
\caption{A short range potential associated with the  Fibonacci chain}
\label{Jane}
\end{figure}

There are several ways to construct a short range potential  associated with the Fibonacci chain. A simple one consists in  
choosing two  real valued smooth functions, $v_{L,L},$  and $v_{S, L}$ with compact support on the interval $(-I, I)$ where 
$0<2I < S (<L)$. A potential  $V_{Fib}$ with range $2L$, can be defined as follows (see Figure \ref{Jane}):
\begin{itemize}
\item for each $n\in \mz$ and for each $\theta \in (s^\infty_n-I, s^\infty_n+I)$:
\begin{itemize}
\item $V_{Fib}(\theta) = v_{L,L}(\theta - s^\infty_n)$ if both intervals $[s^\infty_{n-1}, s^\infty_n]$ and $[s^\infty_n, s^\infty_{n+1}]$ 
have the same length $L$;
\item  $V_{Fib}(\theta) = v_{S,L}(\theta -s^\infty_n)$ if the intervals $[s^\infty_{n-1}, s^\infty_n]$ and $[s^\infty_n, s^\infty_{n+1}]$ 
have different lengths.
\end{itemize}
\item  for $\theta\notin \cup_{n\in \mz} (s'_n-I, s'_n+I),$ $\, V_{Fib}(\theta) = 0$. 
\end{itemize}

The main result of this paper is the following theorem:


\begin{teo}\label{GGP}\footnote{From a more physical point of view, it  is straightforward but interesting to rephrase Theorem \ref{GGP} in terms of  particle density of minimal configurations.}

\noindent
For  the Frenkel-Kontorova model with a  short range potential  associated with a quasicrystal:
\begin{itemize}
\item[(i)] any minimal configuration has a rotation number;
\item [(ii)] the rotation number is a continuous function when defined on the set of minimal configurations  equipped with the product topology;
\item[(iii)] for any $\rho\geq 0$, there exists a minimal configuration with rotation number  $\rho$.
\end{itemize}
\end{teo}

It turns out that, once the appropriate objects have been defined, the proof of Theorem \ref{GGP}  has the same structure as the modern proof for crystals that can be found for instance in \cite{bangaert} or \cite{ le calvez}. More precisely, in the crystal case, a periodic potential factorizes through a  real valued function defined on a circle. In the quasicrystal case, a short range potential factorizes through a real valued function defined on  a more sophisticated compact metric space called the {\it hull }  of the quasicrystal. This hull possesses locally the product structure of  an interval by a Cantor set {\it i.e} it is a solenoid.
This solenoid  can be seen as the suspension of the action of a minimal homeomorphism on the Cantor set. 

\noindent Minimal homeomorphisms on the Cantor set have been extensively studied in topological dynamics and possess a powerful combinatorial description in terms of Kakutani-Rohlin towers (see for instance \cite{GPS}). 
The aim of Section \ref{hull}, which is devoted to the substrate,   is to rephrase these well known results in our specific context, namely for a suspension, in order to see the hull as an inverse limit of one dimensional branched manifolds. These branched manifolds  will play a  central role in the proof. 

\noindent In the crystal case, when projecting a  minimal configuration on the circle, the Aubry-Mather theory shows that it wraps around the circle in a very special  way, namely it is ordered as the orbit of a degree one homeomorphism of circle.  In the quasicrystal case, there exists also some combinatorial obstructions, they are described and analyzed in Section \ref{com} which is devoted to the ground states of the overlying layer.

\noindent Section \ref{preuve} is devoted to the proof of Theorem \ref{GGP}.  First, as for the crystal case, we show, using the inverse limit structure of the hull given in Section \ref{hull} and  the combinatorial obstructions gotten in Section \ref{com},   that minimal configurations have a rotation number (point $(i)$). Then we prove (again as in the crystal case) the continuity of the rotation number (point $(ii)$).  The proof of  point $(iii)$ of Theorem \ref{GGP} in the  crystal case is done first by constructing periodic minimal configurations for any positive rational rotation number and then to use the continuity of the rotation number to get a minimal configuration for any prescribed positive rotation number.
In the quasicrystal case, the scheme is exactly the  same, but the set of rational numbers needs to be replaced by another dense subset of the positive reals. More precisely when the rotation number is not $0$, its  inverse has to be a finite  linear combination with positive integer coefficients of the densities  of patches  of the quasicrystal.

\noindent This paper ends with two final remarks developed in Section \ref{fin}, 
the first one concerns dynamical systems. In the case of a crystal,  minimal configurations for   the Frenkel-Kontorova model are orbits of a twist map on an open annulus. Similarly, in the quasicrystal case, these minimal configurations are also  orbits of a dynamical system that we describe.
The second one consists in  giving the bases of a possible  extension of the theory to quasicrystals in higher dimension.\\

\noindent {\bf Remark:} It should be pointed out that one can find in the literature several studies on the the Frenkel-Kontorova model with a quasi-periodic potential, for instance a potential which is the sum of two periodic potentials with incommensurable periods (see for instance \cite{EFRJ}).  Such potential cannot arise naturally from an underlying one dimensional substrate.  Actually, the underlying object which organizes the minimal configurations and which was a circle in the crystal case and a solenoid in the quasicrystal case, becomes a $2$-torus.  More precisely the real line is immersed as a line with irrational slope in the 2-torus.  Actually, this  is a situation more complex than the one we are dealing with in this paper  which  essentially uses dimension 1 objects, and this  explains the lack of exact results in this quasi-periodic case.\\

\noindent {\it Nota Bene:} To avoid an unnecessary dichotomy and unless explicitly specified, the quasicrystals that appear in the sequel will not be crystals.


\section{The hull of a quasicrystal}\label{hull}
In  this section, we recall some background results concerning quasicrystals. Most of these results  are true  in any dimension  and  they are presented here in the particular case of the dimension 1. Material for Subsections \ref{ds} and \ref{ct} can be found in \cite{KP}, \cite{BBG} and \cite{BG}.  For Subsections \ref{KR} and \ref{invl} a discrete approach can be found in \cite{HPS} and we refer again to \cite{BBG}, \cite{BG} and \cite{Sadun} for a more geometrical point of view. 

\subsection{The hull as a dynamical system}\label{ds}
Consider a chain of atoms whose   configuration is a quasicrystal $\Ss = {(s_n)}_{n\in \mz}$. It is clear that each translated copy $\Ss - u = {(s_n - u)}_{n\in \mz},$  $u\in\mr$, of  $\Ss$ 
is again a quasicrystal. 

\noindent The set of translated copies $\Ss + \mr$ of a quasicrystal can be equipped with a topology  that, roughly speaking, says that two quasicrystal configurations are close one to the other if in a big ball centered at $0$ in $\mr$, the segments of both configurations inside the ball  are equivalent and equal up to a small translation. Such a topology is metrizable and an associated metric can be defined as follows (see \cite{RW} for more details): 

\noindent  Consider two quasicrystal configurations  $\Ss- u_1$ and  $ \Ss-u_2$ in $\Ss + \mr$. Let $A$ denote the set of $\epsilon \in ]0,1[$ for which there exists $u$  with $\vert u\vert < \epsilon,$  such that $\Ss- u_1$ and $\Ss- u_2 + u$ coincide in  $B_{1/\epsilon}(0)$. Then
$$ \delta(\Ss- u_1, \, \Ss-u_2 )\, = \, \inf A \ \ \ \ {\rm if} \ \ A\neq \emptyset$$
$$\ \delta(\Ss- u_1, \, \Ss-u_2) \, =\, 1  \ \ \ \ {\rm if} \ \ A = \emptyset \ .$$

\noindent Hence the diameter of  $\Ss + \mr$ is bounded by $1$ and the $\mr$-action  on $\Ss + \mr$ is continuous. The {\it continuous hull} $\Omega(\Ss)$ of the quasicrystal  $\Ss$ is the completion of the metric space $(\Ss + \mr, \delta)$. 

\noindent As a direct consequence of the {finite local complexity} property, it is easy to check  (see for instance \cite{RW}) that $\Omega(\Ss)$ is a compact metric space and that any element in $\Omega(\Ss)$ is a quasicrystal whose segments are equivalent to segments in $\Ss$.
 The  translation group  $\mr$ acts on $\Omega(\Ss)$  and  the dynamical system $(\Omega(\Ss), \mr)$ possesses (by construction) a dense orbit (namely the orbit $\Ss + \mr$).  On the one hand, the repetitivity property is equivalent to the {\it minimality} of the action {\it i.e}  all its orbits are dense, (see \cite{KP}) and, on the other hand,  the uniform pattern distribution is equivalent to the unique ergodicity {\it {i.e}}  the $\mr$-action possesses a unique  invariant probability measure  (see\cite{BG}).  These results yield  the following proposition.

\begin{prop}\label{min}  Let $\Ss$ be a quasicrystal, then the dynamical system  $(\Omega(\Ss), \mr)$ is minimal  and uniquely ergodic. 
\end{prop}

\noindent In the sequel, we will denote by $\mu$ the unique probability measure on $\Omega(\Ss)$ which is invariant under the $\mr$-action. 

\subsection{The canonical transversal}\label{ct}
The {\it canonical transversal}, $\Omega_0(\Ss),$ of the hull $\Omega(\Ss)$ of a quasicrystal $\Ss$  is the collection of  quasicrystals  in $\Omega(\Ss)$ which contain $0$ ({\it i.e.} such that one atom in the chain is located at  $0$). 
\begin{prop}(see \cite{KP}) 
The canonical transversal of  a quasicrystal  is  either a finite set 
  when $\Ss$ is a crystal  or a Cantor set when not. 
\end{prop}

\noindent It follows that when the quasicrystal $\Ss$ is a crystal,  $\Omega(\Ss)$  is homeomorphic to a circle and when not $\Omega(\Ss)$  has a solenoidal structure, {\it i.e.} it is locally the product of a Cantor set by an interval.

\noindent The {\it return time} function $\Ll: \Omega_0(\Ss) \to \mr^+$  is defined by:
$$\Ll (\Tt) = \inf \{ t  >0\,\,\vert \,\, \Tt \, -\,t \, \in \, \Omega_0(\Ss)\}\quad \forall\,  \Tt\in \Omega_0(\Ss).$$ 

\begin{figure}
\epsfxsize=8truecm
\centerline{\epsfbox{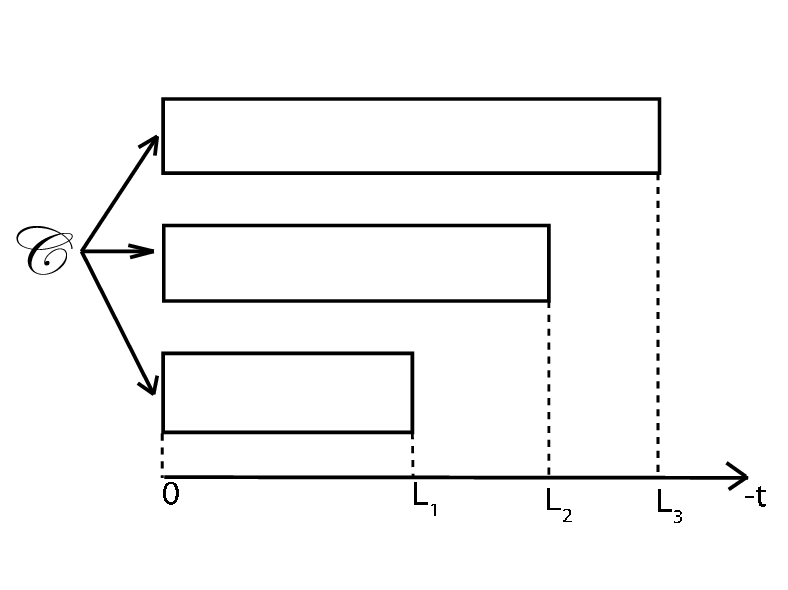}}
\caption{The time return function}
\label{return}
\end{figure}

\noindent The {\it finite local complexity} implies that the function $\Ll$ is locally constant, it  takes finitely many distinct values $L_1, \dots, L_p$ and  the {\it clopen} (closed open)   sets $\Cc_i =\Ll^{-1}(L_i)$ for $i = 1, \dots, p$ form a partition of $\Omega_0(\Ss)$\footnote{ Recall that clopen  sets form a countable basis for the topology of  a totally disconnected  set.} (see Figure \ref{return}).

\noindent The {\it first return map}  $\tau:\Omega_0(\Ss)\to \Omega_0(\Ss)$ is defined by:
$$\tau(\Tt)\, =\, \Tt\, -\, \Ll(\Tt)\quad\forall \, \Tt\,\in\, \Omega(\Ss).$$ The unique invariant probability measure $\mu$ of the $\mr$-action on $\Omega(\Ss)$ induces a finite measure $\nu$  on $\Omega_0(\Ss)$ which is $\tau$-invariant. 

\noindent
For any $i= 1, \dots , p$ and for any clopen set $\Cc$ in $\Cc_i$, the measure $\nu$ satisfies:
$$\nu(\Cc)\,=\, \frac{1}{L_i}\mu(\{(\Tt -  u)\,\quad \Tt\,\in \,\Cc,\quad \, u\, \in \,[0, L_i]\}).$$

\noindent The subsets of $\Omega(\Ss)$ which read $\Cc - u$ where $\Cc$ is a clopen set in one of the $\Cc_i$'s and $u \in [0, L_i[$ are called {\it verticals}.

\noindent The following lemma is a direct byproduct of the above definition:
\begin{lem}\label{chameau} For any $S >0$, there exists a positive constant  $\epsilon_\Ss(S)$ such that, for any vertical $V$ with diameter smaller that $\epsilon_\Ss(S)$ and any pair of configurations $\Ss - x$ and $\Ss - y$ in $V,$  we have:
  $$\Ss\cap B_S(x)\,-\, x =\, \Ss\cap B_S(y) \,  -\, y. $$ 
\end{lem}

 
 \subsection{Potentials on the hull}
The following result shows that a short range potential associated with a quasicrystal $\Ss$,  factorizes through a function on $\Omega(\Ss)$.

\begin{lem}\label{facteur} Let $\Ss$ be a quasicrystal, and let  $V_\Ss$ be a  continuous short range potential associated with $\Ss$. Then, there exists a unique continuous function $\bar V_\Ss: \Omega(\Ss)\to \mr$ such that:
$$ V_\Ss(x) \, =\, \bar V_\Ss(\Ss - x),\quad \forall x\,\in\, \mr.$$ 
Furthermore, when $V_\Ss$ has range $R>0$,  there exists a positive constant  $\epsilon_\Ss(R)$ such that $\bar V_\Ss$ is constant on each vertical with diameter smaller than  $\epsilon_\Ss(R)$.
\end{lem}

\noindent {\bf Remark:} Notice that  when $\Ss$ is a crystal,  Lemma \ref{facteur} simply means that for any continuous periodic function $g: \mr\to \mr$ with period $L$, there exists a continuous function $G: \mr/L.\mz\to \mr$ such that $g = G\circ \pi$, where $\pi: \mr\to  \mr/L.\mz$ is the standard projection. 

\noindent{\it Proof of Lemma \ref{facteur}:}
Assume that $V_\Ss$ is a potential with range $R>0$.  Applying Lemma \ref{chameau}, for any vertical $V$ with   diameter smaller than $\epsilon_\Ss(R)$ and any pair $\Ss-x$ and $\Ss-y$ in $V$,  we have:
  $$\Ss\cap B_R(x)\,-\, x =\, \Ss\cap B_R(y) \,  -\, y, $$  and thus:
$$V_\Ss (x)\,  =\, V_\Ss(y). $$ Since the set $\Ss+ \mr \cap V$ is dense in $V$, it follows that a continuous function $\bar V_\Ss$ which satisfies $ V_\Ss(x) \, =\, \bar V_\Ss(\Ss - x),\,$ $\forall \, x\,\in\, \mr,$  must be constant on $V$ and equal to $V_\Ss(y)$ for any real number  $y$ such that $\Ss - y \in V$. Conversely the function $\bar V_\Ss$ defined this way is clearly continuous, satisfies  $ V_\Ss(x) \, =\, \bar V_\Ss(\Ss - x),\,\forall  \, x\,\in\, \mr,$ and is constant on verticals with diameters smaller than $\epsilon_\Ss(R)$.\hfill$\Box$

\subsection{ Kakutani-Rohlin towers }\label{KR}
The following construction, which has been developed for the study of minimal dynamics on the Cantor set, will be useful all along this paper. It is often referred to as {\it Kakutani-Rohlin towers} (see \cite{HPS}).
Choose $S>0$ and fix a clopen set $\Cc$ in one of the $\Cc_i$'s  with diameter smaller than $\epsilon_\Ss(S)$. 
  
  \noindent Consider the first return time function $\Ll_\Cc$  associated with this clopen set (which is constructed exactly as the first return time function in $\Omega_0(\Ss)$).   The  finite local complexity hypothesis implies that the function $\Ll_\Cc$ is locally constant and   takes finitely many values $L_{\Cc,1}, \dots, L_{\Cc, p(\Cc)}$.  The clopen sets $\D_{\Cc, i} =\Ll_\Cc^{-1}(L_{\Cc, i})$ for $i = 1, \dots, p(\Cc)$ form a partition of $\Cc$.  Again because of the finite local complexity hypothesis, there exists a finite partition of $\Cc$ in clopen sets $\Ee_j$, $j=1, \dots r$ such that for each $j\in\{1, \dots, r\}$,  there exists $i\in \{1, \dots p(\Cc)\}$ so that the following properties are satisfied:
  \begin{itemize}
  \item
  $\Ee_j\subset D_{\Cc, i}$;
   \item for each $ u\in  [0, L_{\Cc,i}[$, $\Ee_j - u$ is a vertical with diameter smaller that $\epsilon_\Ss(S)$.
  \end{itemize}
For $j= 1, \dots, r,$ the  set:
  $$\{\Ee_j \, - \, u, \quad\forall\, u\, \in\, [0, L_{\Cc, i}[\},$$
   is  called a  {\it tower} with {\it height}  $ L_{\Cc, i}$. The union of all these towers realizes a partition of $\Omega(\Ss)$ and the data $(\Ss, S, \Cc,\{ \Ee_j\}_{j\in\{1,\dots, r\}})$ is called a {\it Kakutani-Rohlin towers system with size $S$}.

\begin{figure}
\epsfxsize=8truecm
\centerline{\epsfbox{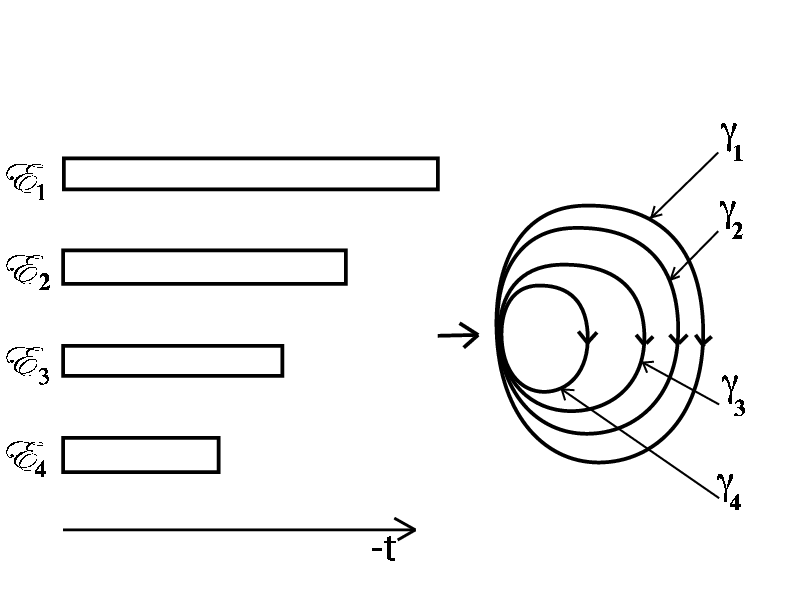}}
\caption{A towers system and its skeleton}
\label{wtc}
\end{figure}

For  $j= 1, \dots, r,$ consider the set  $\Ee_j\subset D_{\Cc, i}$ and 
 for each $ u\in  [0, L_{\Cc,i}[$,  we call {\it floor} of the tower $\Ee_j\times    [0, L_{\Cc,i}[$, the vertical $\Ee_j - u$. By identifying all the points in this vertical, each tower projects on a semi-open interval and the whole hull $\Omega(\Ss)$ projects onto  a smooth  branched one-dimensional manifold  which is  a collection of $r$  of circles $\gamma_1, \dots, \gamma_r$  tangent at a single point.  This branched manifold  is called the  {\it skeleton } of the Kakutani-Rohlin tower system  $(\Ss, S, \Cc,\{ \Ee_j\}_{j\in\{1,\dots, r\}})$.  It  inherits a natural orientation, a differentiable structure and a natural metric  respectively issued from the orientation, the differentiable structure   and the Euclidean metric of the real line $\mr$ (see Figure \ref{wtc}). We denote it $\B$ and call $\pi: \Omega(\Ss)\to \B$ the above identification.
 
\noindent The proof of the following lemma is plain. 
\begin{lem} Let $R>0$ and $V_\Ss$ a continuous potential associated with $\Ss$ with range $R>0$. Consider  a 
 Kakutani-Rohlin towers system   with size $S$  and let $\B$ be its skeleton. 
 
\noindent Assume that $S\geq R$, then the function $\bar V_\Ss:  \Omega(\Ss)\to \mr$ induced by $V_\Ss$ descends to a continuous function $\hat V_\Ss:  \B\to \mr$:
$$\hat V_\Ss\circ \pi\, =\, \bar V_\Ss.$$
Whenever the function $V_\Ss$ is $C^r$-smooth for some $0\leq r\leq\infty$, then the function $\hat V_\Ss$ is also $C^r$-smooth. 
\end{lem}

\subsection{Inverse limits}\label{invl}
Let us  choose an  increasing sequence ${(S_n)}_{n\geq 0}$ going to $+\infty$ with $n$ and let us construct inductively an infinite sequence of Kakutani-Rohlin towers system as follows (see \cite{HPS}):
\begin{itemize}
\item Fix a point $x_0$ in $\Omega_0(\Ss)$.
\item Choose a clopen set $\Cc_0$ containing $x_0$,  with diameter smaller than $\epsilon_\Ss(S_0)$ and construct a  Kakutani-Rohlin towers system $(\Ss, S_0, \Cc_0,\{ \Ee_{0,j}\}_{j\in\{1,\dots, r_0\}})$  with size $S_0$. Up to a renaming of the indices, we can assume that $x_0$ belongs to $\Ee_{0,1}$. We denote by $\B_0$ the corresponding skeleton and  call $\pi_0: \Omega(\Ss)\to \B_0$ the standard projection.
\item We choose a clopen set $\Cc_1\subset \Ee_{0,1}$ which contains $x_0$  with a diameter  small enough so that we can construct a  Kakutani-Rohlin towers system $(\Ss, S_1, \Cc_1,\{ \Ee_{1,j}\}_{j\in\{1,\dots, r_1\}})$  with size $S_1$ such that each of its  towers  intersects all the towers of the previous system.  
Up to a renaming of the indices, we can assume that $x_0$ belongs to $\Ee_{1,1}$. We denote by $\B_1$ the corresponding skeleton and  call $\pi_1: \Omega(\Ss)\to \B_1$ the standard projection.

\item Assume we have constructed a sequence of nested clopen sets $\Cc_n\subset\Cc_{n-1}\subset\dots\Cc_1\subset \Cc_0$ containing $x_0$ and, for each $p =0, \dots, n,$ a Kakutani-Rohlin towers system $(\Ss, S_p, \Cc_p,\{ \Ee_{p,j}\}_{j\in\{1,\dots, r_p\}})$  with size $S_p$ such that each of its  towers  intersects all the towers of the  system associated with $p-1$, and  
such that  $x_0$ belongs to $\Ee_{p,1}$ . We iterate the procedure by choosing  a clopen set $\Cc_{n+1}\subset \Ee_{n,1}$ which contains $x_0$   small enough so that we can construct a  Kakutani-Rohlin towers system $(\Ss, S_{n+1}, \Cc_{n+1},\{ \Ee_{n+1,j}\}_{j\in\{1,\dots, r_{n+1}\}})$  with size $S_{n+1}$  such that each of its  towers  intersects all the towers of the system associated with $n$.  
 Up to a renaming of the indices, we can assume that $x_0$ belongs to $\Ee_{n+1,1}$. We denote by $\B_{n+1}$ the corresponding skeleton and  call $\pi_n: \Omega(\Ss)\to \B_n$ the standard projection.

\end{itemize}
For each $n\geq 0$, fix a point $y$ in $\B_{n+1}$. The set $\pi_{n+1}^{-1}(y)$ is included in a  floor of a tower of the tower system  $(\Ss, S_p, \Cc_p,\{ \Ee_{p,j}\}_{j\in\{1,\dots, r_p\}})$, and thus descends through $\pi_n$  to a single point on $\B_n$. We have defined this way a continuous surjection:
$$\tau_n:\B_{n+1}\to \B_n.$$ 
The inverse limit:
$$\lim_{\leftarrow \tau_n}\B_n\, =\,\{{(x_n)}_{n\geq 0}\, \vert\, x_n \in \B_n\, \,\,{\rm and}\, \,\, \tau_n(x_{n+1})\, =\, x_n,\,\, \forall n\geq 0\}, $$  gives a re-interpretation of the hull $\Omega(\Ss)$:
\begin{prop} \cite {BG} When equipped with the product topology the set $\lim_{\leftarrow \tau_n}\B_n$ is homeomorphic to $\Omega(\Ss)$. 
\end{prop}
Notice that the map $\tau_n:\B_{n+1}\to \B_n$ 
induces  a  $p_n\times p_{n+1}$ {\it homology matrix} $M_n$   whose integer  coefficient $m_{n, i, j}$ is the number of times the  loop $\gamma_{n+1, j}$  in $\B_{n+1}$ covers the loop $\gamma_{n, i}$ of $\B_n$ under the action of the map $\tau_n$.  We remark that the construction of the sequences of towers systems we made insures that, for all $n\geq 0$, the matrix $M_n$ has positive coefficients. These matrices carry information about the invariant measure $\nu$ on the Cantor set through the following lemma (see for instance  \cite{GPS}):
\begin{lem} \label{fabien}
$$ \nu_{n, i}\, =\,\sum_{j=1}^{j=p_{n+1}}m_{n, i, j}\nu_{n+1, j}, \quad \forall i \in \{1, \dots, p(n)\},  $$
where $ \nu_{n, i}$ is the measure of the clopen set  $\Ee_{n, i}$.

\end{lem}

Again the following lemma is plain:
\begin{lem}\label{pierre}
Let $R>0$ and $V_\Ss$ be a continuous potential associated with $\Ss$ with range $R>0$ and choose  an increasing sequence ${(S_n)}_{n\geq 0}$ going to $+\infty$, such that $R\leq S_0$. Then, for each $n\geq 0,$ the  function $V_\Ss$ induces on each branched manifold $\B_n$ a function $\hat V_{\Ss, n}$ which satisfies:
$$\hat V_{\Ss, n}\, \circ \tau_n\, =\, \hat V_{\Ss, n+1}.$$
\end{lem}

\section{Combinatorics of  minimal configurations}\label{com}
\begin{figure}
\epsfxsize=8truecm
\centerline{\epsfbox{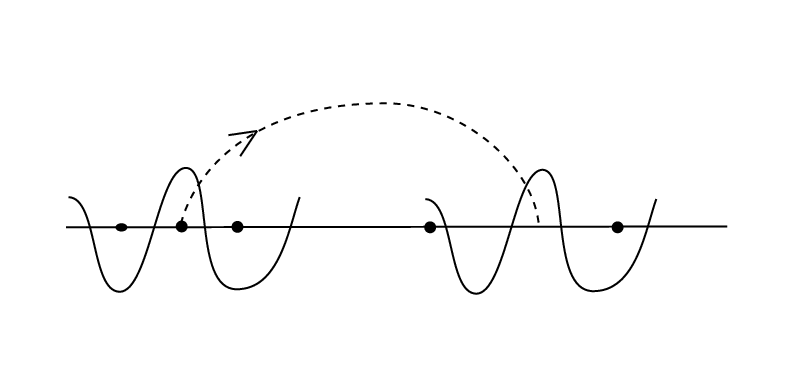}}
\caption{Move of a single atom in  a segment}
\label{ethop}
\end{figure}

In this section, we consider the minimal segments for a short range potential  with range $R$ associated with $\Ss$.
\begin{lem} \label{trou} Let $I$ and $J = I+ u$ be two disjoint intervals in $\mr$ such that for each $\theta$ in $I$:
$$ B_R(\theta)\cap \Ss + u\, =\, B_R(\theta+ u )\cap \Ss,$$
and let $(\theta_1, \dots, \theta_n)$ be a minimal segment  such that $[\theta_1, \theta_n]$ contains $I$ and $J$.  For any pair of  consecutive atoms  $\theta_m$ and $\theta_{m+1}$ in  $I\cap \Ss, $  the interval $[\theta_{m}+ u, \theta_{m+1}+ u]$  contains at most two atoms of the minimal segment.
\end{lem}
\begin{proof}
The proof works by contradiction. Assume that there exists  a pair of atoms   $\theta_m$ and $\theta_{m+1}$ in  $I \cap \Ss, $  such that  the interval $[\theta_{m}+u, \theta_{m+1}+u]$  contains  three consecutive atoms of the minimal segment, say $\theta_l, \, \theta_{l+1}, $ and $\theta_{l+2}$:
$$[\theta_{l}, \theta_{l+2}]\, \subset \, [\theta_{m} +u, \theta_{m+1}+u]. $$ 
We consider the  new segment obtained by taking the atom in position $\theta_{l+1}$ and assigning to it the 
new position $\theta_{l+1} - u$ (Figure \ref{ethop}).
When $u>0$  (what we can assume without loss of generality)  this segment reads:
$$(\theta_1, \dots, \theta_{i}, \dots, \theta_{m}, \theta_{l+1}-u,  \theta_{m+1}, \dots,
\theta_{l}, \theta_{l+2} ,  \dots,  \theta_n).$$

\noindent To get a contradiction we are going to show that the potential energy of this new segment is smaller than the potential energy of the first one.  
On the one hand, since $ B_R(\theta_{l+1})\cap \Ss - u\, =\, B_R(\theta_{l+1}- u )\cap \Ss,$  the potential energy induced by the substrate on the atom that changed its position,  keeps the same value:
$$V_\Ss(\theta_{l+1})\, =\, V_\Ss(\theta_{l+1} \, -\, u).$$
Thus, the sum of the potential energy induced by the substrate on the whole segment is not affected by this change of position. 

\noindent On the other hand, the difference of the potential energy of interaction between the new segment and the former one is given by:
$$\Delta U\, =\, {\large{(}}U(\theta_{m} - \theta_{l+1}+u)\, + U(\theta_{l+1} -u -  \theta_{m+1}) - U (\theta_{m}- \theta_{m+1}){\large{)}} $$ 
$$- {\large{(}}U(\theta_{l} - \theta_{l+1}) + U(\theta_{l+1} - \theta_{l+2})  - U(\theta_{l} -\theta_{l+2}){\large{)}}.$$
Let us introduce the new variables:
$$X\, =\,  \theta_{m} - \theta_{l+1}+u,\quad Y\, =\,\theta_{l+1}-u-  \theta_{m+1},$$
$$X'\, =\, \theta_{l} - \theta_{ l+1},\quad Y'\,=\, \theta_{l+1} - \theta_{l+2}.$$
We have:
$$X\, \leq\,X'\, <\, 0\quad {\rm and}\quad Y\, \leq\, Y'\, <\, 0,$$
 and:
$$\Delta U\, =\, {\large{(}}U(X)\, + U(Y) - U (X + Y){\large{)}} 
- {\large{(}}U(X') + U(Y')  - U(X' + Y'){\large{)}}.$$
For $t\in [0, 1]$, let us consider the function:
$$G(t)\, = \,  U(tX +(1-t)X')\, +\,  U(tY + (1-t)Y')\,- \, U (t(X + Y) + (1-t) (X'+Y')).$$
We have:
$$\Delta U\, =\, G(1)\, -\, G(0),$$
and
\begin{eqnarray*} 
G'(t) & = & 
  U'(tX +(1-t)X')(X - X')\, +\,  U'(tY + (1-t)Y')(Y - Y')\\
& & -\  U' (t(X + Y) + (1-t) (X'+Y'))(X +Y -X' -Y')\\
&=& (U'(tX +(1-t)X')\,- \,U' (t(X + Y) + (1-t) (X'+Y')))(X- X')\\
& & +\  (U'(tY + (1-t)Y') \,-\, U' (t(X + Y) + (1-t) (X'+Y'))) (Y - Y').
\end{eqnarray*}
Observe that  for $t\in [0, 1]$:
$$tX +(1-t)X' \,  \geq\, t(X + Y) + (1-t) (X'+Y')$$
and
$$tY + (1-t)Y'\,\geq\, t(X + Y) + (1-t) (X'+Y').$$
Using the convexity of $U$, more precisely the fact that $U'$ is an increasing function we get that:
$$\Delta U\, \leq\, 0,$$ and this inequality is strict as long as $\theta_m\neq \theta_{l}-u$ and $\theta_{m+1}\neq \theta_{l+2}-u$.
 In this case, we get  the desired contradiction.
 
\noindent In the situation when $\theta_m = \theta_{l}-u$ and $\theta_{m+1} = \theta_{l+2}-u$, we remark that both segments 
$(\theta_m, \theta_{l+1}-u,  \theta-{m+1})$ and $(\theta_{l-1}, \theta_l, \theta_{l+2})$ are not minimal and thus the new 
configuration we constructed is not minimal. The corresponding minimal segment (by fixing the extremities $\theta_1$ and $\theta_n$) 
has an energy which is strictly smaller, a contradiction. 
\end{proof}

The following lemma shows that there are actually more obstructions than the ones described in Lemma \ref{trou}.  

\begin{figure}
\epsfxsize=8truecm
\centerline{\epsfbox{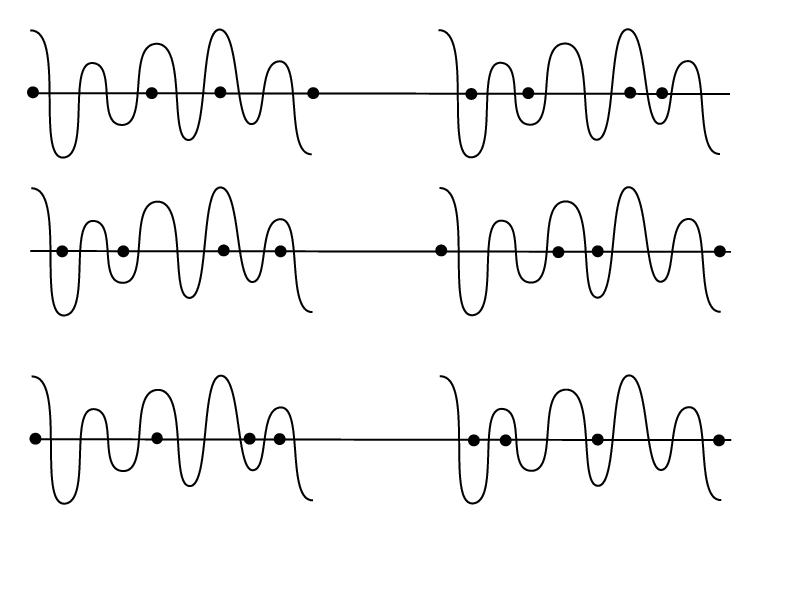}}
\caption{The forbidden 3 situations}
\label{nonnon}
\end{figure}

\begin{lem}\label{pastrou}With the same hypotheses and notations as in Lemma \ref{trou},  consider two disjoint pairs of successive atoms   $\theta_m<\theta_{m+1} <\theta_{m'}<\theta_{m'+1}$ in  $I\cap \Ss, $  such that at least one of the four points $\theta_m +u <\theta_{m+1} +u<\theta_{m'}+u <\theta_{m'+1}+u$ does not belong to the minimal segment.  Concerning  the two intervals  $[ \theta_m +u, \theta_{m+1} +u ]$ and $[\theta_{m'}+u, \theta_{m'+1}+u],$  none of the following  three situations is possible (see Figure \ref{nonnon}):
\begin{itemize}
\item [(i)] both intervals  contain two atoms  of the minimal segment;
\item[(ii)] both intervals  do not contain  atoms  of the minimal segment in their interiors;
\item [(iii)]   one of the interval contains two atoms  of the minimal segment and the other does not contain atoms in its interior.
\end{itemize}

\end{lem}

\begin{proof}  As for Lemma \ref{trou}, we are going to reach a contradiction assuming that situation $(i)$ occurs. The  proof for the other two cases works exactly along the same lines. 
Let  $\theta_l < \theta_{l+1}<\theta_{l'}< \theta_{l'+1}$  be atoms of the minimal segment  such that:
$$[\theta_{l}, \theta_{l+1}]\subset [\theta_{m} +u, \theta_{m+1} +u]$$
and $$
[\theta_{l'}, \theta_{l'+1}]\subset [\theta_{m'} +u, \theta_{m'+1} +u].$$
Assuming  again that $u>0$, let us  move some atoms of the minimal configuration to reach the following new configuration:
$$(\theta_1, \dots, \theta_{m}, \theta_{l+1}-u, \dots, \theta_{l'}-u, \theta_{m'+1}, \dots, 
\theta_{l}, \theta_{m+1}+u, \dots  \theta_{m'}+u,  \theta_{l'+1}, \dots, \theta_n).$$
Since for each $\theta$ in $I$:
$$ B_R(\theta)\cap \Ss + u\, =\, B_R(\theta+ u )\cap \Ss,$$
 the potential energy induced by the substrate on the atoms did not change even if the atoms have changed their positions. 
Thus, the sum of the potential energy induced by the substrate on the whole segment is not affected by this change of position. 

\noindent On the other hand, the difference of the potential energy of interaction between the new segment and the old one is given by:
$$\Delta U\, = \, \Delta U_1 \, + \, \Delta U_2,$$ where
$$ \Delta U_1\, = \, {\large{(}}U(\theta_{m} - \theta_{l+1}+u) + U(\theta_{l} - \theta_{m+1} - u) {\large{)}}
-{\large{(}}U(\theta_{m} -\theta_{m+1}) + U(\theta_{l}- \theta_{l+1}){\large{)}},$$ and

$$ \Delta U_2\, = \, {\large{(}}U(\theta_{l'}  - u - \theta_{m'+1}) + U(\theta_{m'}+u - \theta_{l'+1} ) {\large{)}}
-{\large{(}}U(\theta_{m'} -\theta_{m'+1}) + U(\theta_{l'}- \theta_{l'+1}){\large{)}}.$$
Let us introduce the new variables:
$$X_0 \, = \, \theta_{m}  \,\quad \, X_1\,=\,  \theta_{l+1 } -u  \quad {\rm and}\quad  Y_0\, =\, \theta_{l} -u\quad Y_1\, =\, \theta_{m+1}.$$ 
We have:
$$\Delta U_1\, =\,  \, {\large{(}}U(X_0 - X_1) + U(Y_0 -Y_1) {\large{)}}
-{\large{(}}U(X_0-Y_1) + U(Y_0 - X_1){\large{)}}.$$
This yields:
$$\Delta U_1\, =\,  \,-\, \int_{X_0}^{Y_0}\left(\int_{X_1}^{Y_1} U''(v - u) du\right)\,dv.$$
Since $U$ is convex,   $X_0\leq Y_0$ and $X_1\leq Y_1$ and at least one of these inequalities is strict,  we get:
$$\Delta U_1\, <\, 0, $$ and for the same reason
$$\Delta U_2\, <\, 0. $$
This yields a contradiction.
\end{proof}

From the  previous two lemmas, we deduce that  the quantity of atoms of the minimal segments which belong to $I$ and to $I+u$ differ by an integer smaller than 2. This is summarized in the following proposition that will be our main tool in the sequel of this paper.
\begin{prop}\label{borne}
Let $(\theta_1, \dots, \theta_n)$ be a minimal segment and let  $I$ be an interval in  $[\theta_1, \theta_n]$, then there exists an integer $N\in \mz^+ $ such that for any pair of disjoint intervals $I_1 =  I+u_1$ and $ I_2= I+u_2$   in $[\theta_1, \theta_n]$ which satisfy that for each $\theta$ in $I$ and $k=1,\, 2$:
$$ B_R(\theta)\cap \Ss + u_k\, =\, B_R(\theta+ u_k )\cap \Ss,$$
each interval $I_k$ contains either $N$,  $N+1$ or $N+2$   atoms of the minimal segment.
\end{prop}
\section{Proof of Theorem \ref{GGP}} \label{preuve}
\subsection{Existence of a rotation number}
In this subsection, we consider a minimal configuration  for a  potential with range $R$ associated with $\Ss$.
\noindent
Let us consider an increasing sequence ${(S_l)}_{l\geq 0}$ going to $+\infty$ with $l$ and such that $S_0>R$ and consider also  an associated sequence of  Kakutani-Rohlin towers systems   ${(\Ss, S_l, \Cc_l,\{ \Ee_{l,j}\}_{j\in\{1,\dots, r_l\}})}_{l\geq 0}$  and the corresponding sequence of skeletons ${(\B_n)}_{n\geq 0}$ as constructed in Subsection \ref{invl}. 

\noindent The identification $$I:x\in\mr\mapsto \Ss - x\in \Omega(\Ss)$$ induces an immersion of the real line in $\Ss$ and the image of a configuration ${(\theta_n)}_n$ through this immersion is an element ${(\bar\theta_n)}_n$ in $\Omega (\Ss)^\mz$ where $\bar\theta_n = \Ss -\theta_n,$ for all $n\in \mz$ . In turn,  for any $l\geq 0,$ the projection $\pi_l: \Omega(X)\to \B_l$ transforms this sequence in an element ${(\hat\theta^l_n)}_n$  in $\B_l^\mz$ where $\hat\theta^l_n = \pi_l (\bar\theta_n) = \pi_l\circ I (\theta_n),$ for all $n\in \mz$. Furthermore we have:
$$\hat V_{\Ss, l}(\hat\theta_n) \, =\, \bar V_\Ss(\bar\theta_n)\, =\, V_\Ss(\theta_n).$$

\noindent The following lemma is a direct consequence of Proposition \ref{borne}:
\begin{lem} \label{turc}
Let   ${(\theta_n)}_n$ be a minimal configuration such that $\lim_{n\to +\infty} \theta_n =+\infty$ and  $\lim_{n\to -\infty} \theta_n =-\infty$ (resp. let $(\theta_p, \dots, \theta_q)$ be a minimal segment).  Then, for any $l\geq 0$ and any $ j\in \{1, \dots, r_l\}$, there exists an integer $N_{l, j}$ such that for each loop $\gamma_{l, j}$ of $\B_l$, each connected component  of $(\pi_l\circ I)^{-1}(\gamma_{l, j})\subset \mr$  (resp. each connected component  of $(\pi_l\circ I)^{-1}(\gamma_{l, j})\subset \mr$  which does not intersect $(-\infty, \theta_p]\cup [\theta_q, +\infty)$) contains either $N_{l,j}$ or $ N_{l,j}+1$ or $N_{l, j}+2$ atoms of the minimal configuration (resp. the minimal segment).  

\noindent In other words, when $n$ increases, the projection of the minimal configuration (resp. the  minimal segment) stays the same amount of time in a given loop up to an error of $2$. 
\end{lem}

Now we can prove the existence of a non negative rotation number for any minimal configurations.
 
 First, consider a minimal configuration  ${(\theta_n)}_n$ such that $\lim_{n\to +\infty} \theta_n =+\infty$ and  
$\lim_{n\to -\infty} \theta_n =-\infty$.  Let us estimate the length of the interval $[\theta_0, \theta_n]$ for $n\geq 0$.  
Let $n_{l, j}$ be the number of times $\pi_l\circ I ([\theta_0, \theta_n])$ covers completely the loop  
$\gamma_{l, j}$ of $\B_l$. We have, for each $l\geq 0$:

$$ \sum_{j=1}^{p_l} n_{l,j} L_{l, j} \, \leq\,   \theta_n -\theta_0\,\leq\,   \sum_{j=1}^{p_l} n_{l,j} L_{l, j} \, +\, 2 L_l, $$
where $L_{l, j}$ is the height of the tower associated with the loop $\gamma_{l, j}$ ({\it i.e.} the length of the loop $\gamma_{l, j}$)  and 
$$L_l\, =\, \max_{j\in \{1, \dots,p_l\}} L_{l, j}.$$
On the other hand we have:

$$ \sum_{j=1}^{p_l} n_{l,j}N_{l, j}\, \leq \,n \, \leq \,  \sum_{j=1}^{p_l} n_{l,j}(N_{l,j} +2)\,+\, 2(N_l +2),$$
where 
$$N_l\, =\, \max_{j\in \{1, \dots,p_ l\}} N_{l, j}.$$
This yields:
$$ \frac{\sum\limits_{j=1}^{p_l} n_{l,j} L_{l, j}}{ \sum\limits_{j=1}^{p_l} n_{l,j}(N_{l,j} +2)\,+\, 2(N_l +2)} \, \leq\,   \frac{\theta_n -\theta_0}{n}\,\leq\,   \frac{\sum\limits_{j=1}^{p_l} n_{l,j} L_{l, j} \, +\, 2 L_l}{ \sum\limits_{j=1}^{p_l}n_{l,j}N_{l, j}}.$$ 
When $n$ goes to $+\infty$ the quantity:
$$ \frac{n_{l,j}}{\sum\limits_{j=1}^{p_l} n_{l,j} L_{l, j}}$$ goes to the measure $\nu_{l, j}$ of the clopen set $E_{l, j}$.
It follows that the sequence $(\theta_n- \theta_0)/n$ has bounded limit sup and limit inf and that any  accumulation point  $\rho$ of this sequence  satisfies:
$$ \frac{\sum\limits_{j=1}^{p_l}\nu_{l,j} L_{l, j}}{ \sum\limits_{j=1}^{p_l} \nu_{l,j}(N_{l,j} +2)} \, \leq\,  \rho\,\leq\,   
\frac{\sum\limits_{j=1}^{p_l} \nu_{l,j} L_{l, j} }{ \sum\limits_{j=1}^{p_l}\nu_{l,j}N_{l, j}}.$$ 
Recall that the measure $\nu$ is the transverse measure associated with an invariant  probability measure on the hull $\Omega(\Ss)$ and thus:
$$\sum_{j=1}^{p_l} \nu_{l,j} L_{l, j}\, =\, 1.$$ On the other hand we have:
$$\sum_{j=1}^{p_l} \nu_{l,j}\, =\, \nu(\Cc_l).$$
We deduce that:
$$ \frac{1}{\sum\limits_{j=1}^{p_l} \nu_{l,j}N_{l,j} +2\nu(\Cc_l)} \, \leq\,  \rho\,\leq\,   \frac{1}{\sum\limits_{j=1}^{p_l}\nu_{l,j}
N_{l, j}}.$$ 
Since these last inequalities are true for any $l\geq 0$, and since $\nu(\Cc_l)$ goes to $0$ as $l$ goes to $+\infty$,  it follows that the sequence  $(\theta_n- \theta_0)/n$ converges to the limit:

\[
\lim_{l\to +\infty}  \frac{1}{ \sum\limits_{j=1}^{p_l}\nu_{l,j}N_{l, j}} \qquad (\star).
\]
Observe that this rotation number is different from $0$.

Consider now a minimal configuration which  satisfies $\lim_{n\to +\infty} \theta_n = M<+\infty$  or $\lim_{n\to +\infty} \theta_n = m<+\infty$.
 The constant configuration
$$\theta_n \, =\, \theta_0, \quad \forall n\in \mz,$$  has obviously a rotation number  equal to $0$. Let us assume now that the minimal configuration is not constant and satisfies $\lim_{n\to +\infty} \theta_n = M<+\infty$. Let us show that we cannot have $\lim_{n\to -\infty} \theta_n = -\infty$. Indeed, consider the interval $[M -2R, M+2R]$ and choose $u>0$ such that the interval  $[M -2R - u , M+2R- u]$ is disjoint from $[M -2R, M+2R]$ and such that:
$$B_{2R}(M-u)\cap \Ss \, +\, u\, =\, B_{2R}(M)\cap \Ss.$$
Consider now, for $n$ large enough, the interval $[\theta_n -R, \theta_n] \subset [M -2R, M]. $ The number of atoms in $[\theta_n -R, \theta_n] $ goes to $+\infty$ with $n$. If 
 $\lim_{n\to -\infty} \theta_n = -\infty$, it follows from Proposition \ref{borne} that the number of atoms in $[\theta_n -R - u, \theta_n -u] $ and thus in  $[M -2R - u , M+ 2R - u]$,    goes to $+\infty$ with $n$. Consequently the minimal sequence ${(\theta_n)}_n$ has an accumulation point in $[M -2R - u , M+ 2R - u]$ when $n$ goes to $-\infty$ which is a contradiction.
Thus for a minimal configuration we have:

\begin{eqnarray*}
\lim_{n\to +\infty} \theta_n <+\infty &\Longleftrightarrow & \lim_{n\to -\infty} \theta_n >- \infty\\
& \Longleftrightarrow &  {(\theta_n)}_n \quad \text{ is bounded }\\
& \Longleftrightarrow &  {(\theta_n)}_n \quad \text{ has rotation number } 0.
\end{eqnarray*}

This ends the proof of Part $(i)$ of Theorem \ref{GGP}.

\subsection{Continuity of the rotation number}
Consider a sequence   ${(\theta_{m,  n})}_n$ of minimal configurations with rotation numbers $\rho_m$ which converges,  in the product topology, to a minimal configuration   ${(\theta_ n)}_n$  with rotation number $\rho> 0$.  We fix $l>0$ and   choose a loop $\gamma_{l, j}$  in $\B_{l}$. Consider the first time when, starting from $0$ on the real line and going in the positive direction, the projection of the configuration  ${(\theta_n)}_n$ enters in this loop.  Let us do the same for the configuration  ${(\theta_{m, n})}_n$. Since    ${(\theta_{m,  n})}_n$ converge to    ${(\theta_n)}_n$ in the product topology, for $m$ large enough both configurations stay the same time in the loop for their first visits. It follows from Lemma \ref{turc}  that  the minimal number   of times $N_{m, l, j}$, the  projections of the configurations ${(\theta_{m, n})}_n$ spend in the loop  $\gamma_{l, j}$ of $\B_{l}$,   and  the minimal number  of times  $N_{ l, j}$,  the projection of the configuration ${(\theta_{m, n})}_n$ spends in the same  loop $\gamma_{l, j}$,  satisfy:
$$\vert N_{m, l, j}\,-\,  N_{ l, j}\vert \, \leq \, 2,\quad \forall  \, j\in \{1, \dots, p(l)\}.$$
The rotation  number  $\rho_m$ of the configuration  ${(\theta_{m, n})}_n$  satisfies:
$$\frac{1 }{ \sum\limits_{j=1}^{p_{l}}\nu_{l,j}N_{m, l, j} + 2\nu(\Cc_{l})}\,\leq\, \rho_m\, \leq \, 
\frac{1 }{ \sum\limits_{j=1}^{p_{l}}\nu_{l,j}N_{m, l, j} }.$$
On the other hand 
$$\frac{1}{ \sum\limits_{j=1}^{p_{l}}\nu_{l,j}N_{l, j} + 2\nu(\Cc_{l})}\,\leq\, \rho\, \leq \, 
\frac{1}{\sum\limits_{j=1}^{p_{l}}\nu_{l,j}N_{ l, j} }.$$
This implies that for $m$ large enough:
$$\left\vert \frac{1}{\rho}\, -\, \frac{1}{\rho_m}\right\vert \, \leq\, 8\nu(\Cc_{l}).$$
Considering bigger and bigger $l$ yields:
$$\lim_{m \to +\infty} \rho_m\, =\, \rho.$$
When the rotation number $\rho =0$, we have proved that the configuration  ${(\theta_n)}_n$ is bounded. Let $M$ be its upper bound and  consider the loop $\gamma_{0,i}$  in $\B_0$ on which $M$ descends by projection.  If $M$ falls on the singular point, we consider the loop where  the $M-\epsilon$'s  for $\epsilon>0$ small enough, are falling. Fix $ K>2$, when $m$ is big enough, the projection of the configuration    ${(\theta_{m, n})}_n$ (whose rotation number is assumed to be different from $0$)  must spend at least a time $K$ in the loop  $\gamma_{0,i}$ during one of its visits and thus, thanks to Lemma \ref{turc}  at least $K-2$ times at each of its visits. It follows that the rotation number of   ${(\theta_{m, n})}_n$ satisfies:
$$\rho_m\, \leq\,  \frac{1}{(K-2) \nu_{0, i}},$$  and, consequently:
$$\lim_{m \to +\infty} \rho_m\, =\, 0.$$
Thus, we have proved Part $(ii)$ of Theorem \ref{GGP}. 

\subsection{Construction of minimal configurations}
Observe that a constant configuration is a minimal configuration with rotation number $0$. 
For positive rotation numbers,  we are first going 
to construct minimal configurations for a dense subset of rotation numbers in $\mr^+$.

The good candidate $\Ff$  to be a dense set in $\mr^+$ for which minimal configurations can be construct is suggested  
by the expression $(\star)$  obtained  in the previous subsection. Again, let us consider an increasing sequence 
${(S_l)}_{l\geq 0}$ going to $+\infty$ with $l$ and such that $S_0>R$. Consider also  an associated sequence of  
Kakutani-Rohlin towers systems  ${(\Ss, S_l, \Cc_l,\{\Ee_{l,j}\}_{j\in\{1,\dots, r_l\}})}_{l\geq 0}$  and the corresponding 
sequence of skeletons ${(\B_n)}_{n\geq 0}$ as constructed in Subsection \ref{invl}.  Recalling that the  $\nu_{l,j}$'s are 
the measures of the clopen sets $\Ee_{l,j}$, we define the set $\Ff$ as follows:

 $$\Ff\, =\, \left\{\frac {1}{\sum\limits_{j=1}^{p_l}N_{l,j}\nu_{l,j}}, \quad 
\forall N_{l,j}\in \mz^+ \setminus\{0\} , \quad \forall j\in\{1, \dots , p_l\}, \quad \forall l\geq 0\right\}.$$
Since the measures of the clopen sets $\Ee_{l,j}$ go to zero with $l$ uniformly in $j$, we check easily that 
$\Ff$ is a dense subset of $\mr^+$.

\begin{prop}\label{18} For any real number  $\rho_0$ in $\Ff$, there exists a minimal configuration 
with rotation number $\rho_0$.
\end{prop}
\begin{proof}
 Fix $ l_0\geq 0$ and choose  $p_{l_0}$ positive integers  $N_{l_0, 1},\dots,N_{l_0, p_{l_0}}$. Consider the positive real number:
$$\rho_0\, =\, \frac {1}{\sum\limits_{j=1}^{p_{l_0}} N_{l_0,j}\nu_{l_0,j}}\, \in\, \Ff.$$
Let us construct  a minimal configuration with rotation number $\rho_0$.

\noindent {\bf Step 1:} For $j=1, \dots , p_{l_0}$, consider on the  loop  $\gamma_{l_0, j}$ of the oriented 
branched manifold $\B_{l_0}$, $N_{{l_0}, j} -1$ points $ \hat b_{{l_0}, 1} <\dots < \hat b_{ {l_0}, N_{{l_0}, j} -1}, $  
disjoint from the singular point $\pi_{l_0}(x_0)$ of $\B_{l_0}$ 
(where we recall that $\cap_{l\geq 0}\  \Cc_l\, =\, \{x_0\}$) as shown in Figure \ref{mark}.
\begin{figure}
\epsfxsize=6truecm
\centerline{\epsfbox{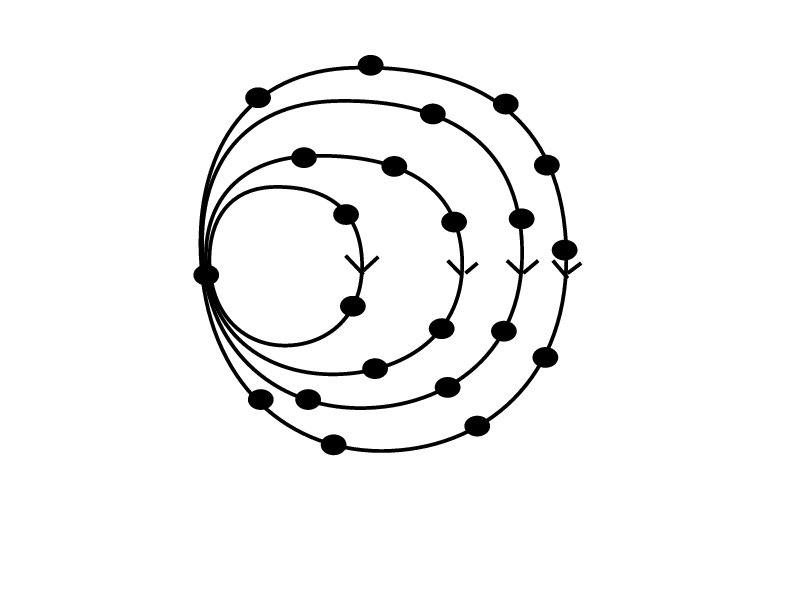}}
\caption{The branched manifold with its marked points}
\label{mark}
\end{figure}
For each $j =1, \dots, p_{l_0},$ we consider the segment:
$$(\pi_{l_0}(x_0) ,  \hat b_{{l_0}, 1} , \dots , \hat b_{ {l_0}, N_{{l_0}, j} -1}, \pi_{l_0}(x_0) ).$$
Thanks to Lemma  \ref{pierre}, it makes sense to compute the potential energy of this segment and to consider the position of the points, $\hat b_{{l_0}, 1} , \dots , \hat b_{ {l_0}, N_{{l_0}, j} -1}, $ which minimizes this potential energy.  Having done it for all loops, we  denote $\hat B_{l_0}$  the collection of these marked points (all the $\hat b_{{l_0}, k}$'s and $\pi_{l_0}(x_0)$)  on $\B_{l_0}$ and consider the  subset of the real line $(\pi_{l_0}\circ I)^{-1} (\hat B_{l_0})$. It is a discrete subset that we can ordered as a bi-infinite increasing sequence ${(\theta_{l_0,n})}_n$. This subset  of $\mr$ contains  the subset  $(\pi_{l_0}\circ I)^{-1} (\pi_{l_0}(x_0))$  which is a quasicrystal.  The configuration ${(\theta_{l_0,n})}_n$ is made with a concatenation of minimal segments whose extremities are consecutive points in  $(\pi_{l_0}\circ I)^{-1} (\pi_{l_0}(x_0))$ and,  there are exactly $p_{l_0}$ different equivalence classes of segments, each of them corresponding to a minimal segment starting at the beginning and ending at the end of a loop in $\B_{l_0}$.

\noindent {\bf Step 2:}  Consider now the subset $\tau_{l_0}^{-1}(\hat B_{l_0})$ of the branched manifold $\B_{l_0+1}$.  This subset contains the singular point $\pi_{l_0+1}(x_0)$  and for each $j= 1, \dots, p_{l_0+1},$ the  loop $\gamma_{l_0+1, j}$  of $\B_{l_0+1}$ contains
$N_{l_0+1, j} -1$  consecutive points, $\hat b_{l_0+1, 1}<\dots  <\hat b_{l_0+1,N_{l_0+1, j} -1,  } $ distinct from the singular point $\pi_{l_0+1}(x_0)$.  
Actually we have:
$$ N_{l_0+1, j}\,=\, \sum\limits_{i=1}^{p_{l_0}} m_{l_0, i, j} N_{{l_0}, i} \qquad (\star \star)$$ 
where $m_{l_0, i, j}$ is the coefficient of the homology matrix $M_{l_0}$.
 Again, for each $j =1, \dots, p_{l_0+1},$ we consider the segment:
 $$(\pi_{l_0+1}(x_0) ,  \hat b_{{l_0+1}, 1} , \dots , \hat b_{ {l_0+1}, N_{{l_0+1}, j} -1}, \pi_{l_0+1}(x_0) ).$$
we  choose the position of the points $\hat b_{{l_0+1}, 1} , \dots , \hat b_{ {l_0+1}, N_{{l_0+1}, j} -1} $ which minimizes the potential energy.  Having done it for all loops, we  denote $\hat B_{l_0+1}$  the collection of these marked points (all the $\hat b_{{l_0+1}, k}$'s and $\pi_{l_0+1}(x_0)$)  on $\B_{l_0+1}$ and consider the  subset of the real line $(\pi_{l_0+1}\circ I)^{-1} (\hat B_{l_0+1})$. It is a discrete subset that we can ordered as a bi-infinite increasing sequence ${(\theta_{l_0+1,n})}_n$.
This subset  of $\mr$ contains  the subset  $(\pi_{l_0+1}\circ I)^{-1} (\pi_{l_0}(x_0))$  which is a quasicrystal contained in the quasicrystal $(\pi_{l_0}\circ I)^{-1} (\pi_{l_0}(x_0))$.  The configuration ${(\theta_{l_0+1,n})}_n$ is made with a concatenation of minimal segments whose extremities are consecutive points in  $(\pi_{l_0+1}\circ I)^{-1} (\pi_{l_0+1}(x_0))$ and there are exactly $p_{l_0+1}$ equivalence classes of segments, each of them corresponding to a minimal segment starting at the beginning and ending at the end of a loop 
in $\B_{l_0+1}$ (See Figure \ref{fkbis}).
\begin{figure}
\epsfxsize=8truecm
\centerline{\epsfbox{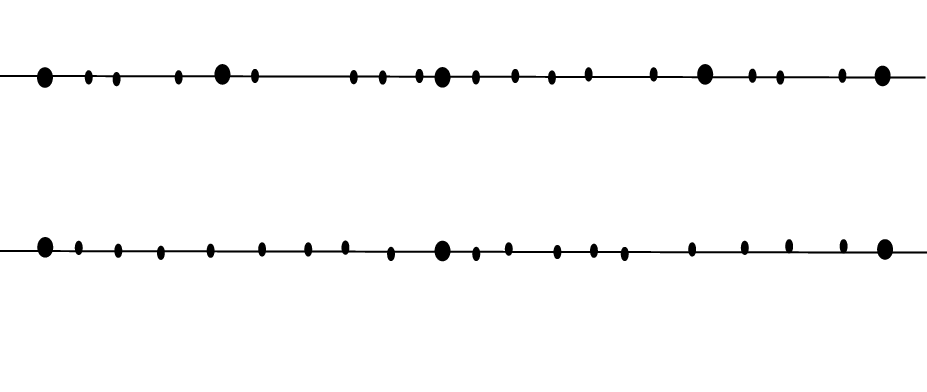}}
\caption{The configurations  ${(\theta_{l_0,n})}_n$ and  ${(\theta_{l_0+1,n})}_n$}
\label{fkbis}
\end{figure}

\noindent {\bf Step 3:}   We iterate this procedure to get a configuration  ${(\theta_{l_0+m,n})}_n$ for each $m\geq 0$.

\begin{lem} For each $m\geq 0$, the configuration  ${(\theta_{l_0+m,n})}_n$ has  rotation number $\rho_0$.
\end{lem}

\begin{proof} 
As a preliminary remark, observe that by construction:
\begin{itemize}
\item For any $j$ in $\{1, \dots, p(l_0)\}$, each time the projection of   the configuration ${(\theta_{l_0,n})}_n$  crosses  
the  loop $\gamma_{l_0, j}$  of $\B_{l_0}$, it spends an amount of time $N_{l_0, j}$ in this loop. 

\item Similarly, for any $m\geq 0$ and for any $k$ in $\{1, \dots, p(l_0+m)\}$, each time the projection of the configuration 
${(\theta_{l_0+m,n})}_n$ crosses the loop $\gamma_{l_0+m, k}$  of $\B_{l_0+m}$, it spends an amount of time 
$N_{l_0+m, k}$ in loop. 

\item Remark also that for any $m\geq 0$ and any  $k$ in $\{1, \dots, p(l_0+m)\}$, each time the projection of the configuration 
${(\theta_{l_0,n})}_n$  crosses  the  loop  $\gamma_{l_0+m, k}$  of $\B_{l_0+m}$, it spends an amount of time  
in loop which  is precisely  $N_{l_0+m, k}$.

\end{itemize}

Using the same estimate as for the proof of the existence of a rotation number for a minimal configuration, we get that 
the configuration  ${(\theta_{l_0,n})}_n$ has a rotation number and that this rotation number is the limit 
when $m\to +\infty$ of the sequence ${(\rho_m)}_{m\geq 0}$, where:
$$\rho_m\, =\, \frac {1} {\sum\limits_{j=1}^{p_{l_0+m}}\nu_{l_0+m,j}N_{l_0+m, j}}\quad \forall \, {m\geq 0}.$$

\noindent {\bf  Claim:} {\it The sequence ${(\rho_m)}_{m\geq 0}$ is constant.}

\noindent {\it Proof of the claim:}
Using the relation $(\star\star)$ we get, for each $m\geq 0$:

\begin{eqnarray*}
\frac {1}{\sum\limits_{j=1}^{p_{l_0+m+1}}\nu_{l_0+m+1, j}N_{l_0+m+1, j}}
& = & 
\frac {1}{\sum\limits_{j=1}^{p_{l_0+m+1}}\nu_{l_0+m+1, j}\left(\sum\limits_{i=1}^{p_{l_0+m}} m_{l_0+m, i, j} N_{{l_0+m}, i}\right)}\\
& = &
\frac{1}{\sum\limits_{i=1}^{p_{l_0+m}} N_{{l_0+m}, i}\left(\sum\limits_{j=1}^{p_{l_0+m+1}}m_{l_0+m+1, i, j}
\nu_{l_0+m+1, j}\right)}.
\end{eqnarray*}
%
Thanks to   Lemma \ref{fabien}:
$$\nu_{l_0+m, i}\, =\,\sum_{j=1}^{p_{l_0+m+1}}m_{l_0+m+1, i, j}\nu_{l_0+m+1, j}.$$
Thus:
$$\rho_{m+1}\, =\, \rho_m, \quad \forall \, m\geq 0.$$
This proves the claim and shows that the rotation number of the configuration ${(\theta_{l_0, n})}_n$ is equal to $\rho_0$.

 To conclude the proof of the lemma, we remark  that a same computation yields that, for each $p\geq 0$, the configuration  
${(\theta_{l_0+p,n})}_n$ has a rotation number and that this rotation number is the limit when $m\to +\infty$ of the sequence 
${(\rho_{p,m})}_{m\geq 0}$, where:
$$\rho_{p, m}\, =\, \frac {1}{\sum\limits_{j=1}^{p_{l_0+p+m}}\nu_{l_0+p+m,i}N_{l_0+p+m, j}}\quad \forall \, {m\geq 0}.$$ 
As shown previously, the sequence  ${(\rho_{p,m})}_{m\geq 0}$ is constant and  $\rho_{p,0} = \rho_p = \rho_0$. 
\end{proof}

\noindent {\bf Step 4:}  
\begin{lem}\label{grec} There exists  $M>0$ such that:
$$0\leq \theta_{l_0+ m, n+1}\, -\, \theta_{l_0+m, n}\, \leq \, M\quad\quad  \forall \, m\geq 0, \quad \forall\, n\in \mz.$$
\end{lem}
\begin{proof}  Notice first that because of the very construction of the configurations ${(\theta_{l_0+m, n})}_n$ 
the lemma is true if we consider only a finite subset of these sequences. 
Let us prove this lemma by contradiction. Let us fix $m_0>0$ and assume that the lemma is not true for the set of sequences  
${(\theta_{l_0+m, n})}_n$ with $m>m_0$.  Choose $M(m_0)>0$ such that $M(m_0)$ is larger  than the longest loop of $\B_{l_0+m_0}$.  
We know that there exists $m>m_0$ and $n\in \mz$ such that:$$M(m_0) < \theta_{m, n+1}\, -\, \theta_{m, n}.$$ Recall that the 
configuration ${(\theta_{l_0+m, n})}_n$ is a concatenation of minimal segments whose extremities descend by projection 
on the singular point of $\B_{l_0+m_0}$. This implies that there exists a minimal segment:
$$\Theta\, =\, (\theta_{l_0+m, n_1}, \dots, \theta_{l_0+m, n},  \theta_{l_0+m, n+1},\dots,  \theta_{l_0+m, n_2})$$ 
of the configuration ${(\theta_{l_0+m, n})}_n$  and a loop $\gamma_{l_0+m, j}$ in $\B_{l_0+m_0}$  such that:
$$\pi_{l_0+m_0}\circ I (\Theta)\cap\gamma_{l_0+m_0, j}\, =\, \emptyset, \quad {\rm and}\quad  \gamma_{l_0+m_0, j}\,
\subset\, \pi_{l_0+m_0}\circ I ([ \theta_{l_0+m, n},  \theta_{l_0+m, n+1}]).$$ 
Recall that the image $\tau_{l_0+m_0-1}(\gamma_{l_0+m_0, j})$ covers all the loops of  $\B_{l_0+m_0-1}$.   
Using Lemma \ref{turc}, we deduce that  the projection of the segment  $\Theta$  on $\B_{l_0+m_0 -1}$ stays at 
each passage in a loop 	 of $\B_{l_0+m_0 -1}$ at most  3 times in this loop.
It follows that the rotation  number  $\rho_{l_0+m}$ of the configuration ${(\theta_{l_0+m, n})}_n$  satisfies:

$$\rho_{l_0+m}\,\geq \, \frac{1}{3\nu(\Cc_{l_0+m_0-1})}.$$
This inequality must be true for all $m_0\geq 0$  and thus $\rho_0 = +\infty$, a contradiction. 
\end{proof}

Let us consider the set $\mr^\mz$ equipped with the product topology.  
For $M>0$, the set  $S_M$ of non decreasing sequences ${(\xi_n)}_n$ in $\mr^\mz$ such that:
$$0\leq  \xi_n -\xi_{n-1}\, \leq M, \quad \forall n\in \mz,$$ is a compact subset of $\mr^\mz$. Thus it follows from Lemma \ref{grec}, 
that the set of all the configurations  ${(\theta_{l_0+m, n})}_n$, for $m\geq 0$  and  their translated is 
in a compact subset of $\mr^\mz$. 

\noindent {\bf Step 5:}  
For each $m\geq 0$, consider $u_m\in \mr$ such that $0$ belongs to the center of a minimal segment of ${(\theta_{l_0+m, n} +u_m)}_n$. 
From lemma \ref{turc}, the sequence of configurations ${(\theta_{l_0+m, n}+u_m)}_n$ has an accumulation point in 
$\mr^\mz$. We denote this configuration ${(\theta_{\infty, n})}_n$.

\begin{lem}\label{bad} 
The configuration ${(\theta_{\infty, n})}_n$ is a minimal configuration with rotation number $\rho_0$.  
\end{lem} 
\begin{proof} The fact that the configuration  ${(\theta_{\infty, n})}_n$ is minimal is standard. Consider a segment of  
${(\theta_{\infty, n})}_n$. By construction this segment is a limit of minimal segments and it is straightforward to 
show that this segment is minimal. 

\noindent
Let  us prove now that the configuration  ${(\theta_{\infty, n})}_n$ has rotation number $\rho_0$.  
Since the configuration is minimal, it has a rotation number $\rho_\infty$ which is defined as the limit:
$$\lim_{l\to +\infty}  \frac{1 }{ \sum\limits_{j=1}^{p_l}\nu_{l,j}N_{\infty, l, j}},$$ 
where $N_{\infty, l, j}$ is 
the minimal number of times the configuration ${(\theta_{\infty, n})}_n$ spends in the $j^{th}$ loop of $\B_l$.

\noindent We use a similar argument to the one used in the proof of the continuity of the rotation number.  
Fix $l_1 >l_0$,  and  choose a loop $\gamma_{l_1, j}$  in $\B_{l_1}$. Consider the first time when, starting from $0$ 
on the real line and going in the positive direction, the configuration ${(\theta_{\infty, n})}_n$ enters in this loop.  
Let us do the same for the configuration  ${(\theta_{l_0+m, n}+u_m)}_n$.  Since a subsequence of configurations 
${(\theta_{l_0+m, n} +u_m)}_n$ converges, when $m$ goes $+\infty$ to  the configuration 
${(\theta_{\infty, n})}_n$, it follows that  for $m$ big enough both projections of the configurations stay the same 
time in the loop  $\gamma_{l_1, j}$  for their first visit in this loop. It follows from Lemma \ref{turc}  
that  the minimal number  $N_{l_0+m, l, j}$  of times the projection of the configuration ${(\theta_{l_0+m, n})}_n$ spends 
in the loop $\gamma_{l_1, j}$  of $\B_{l_1}$ satisfies:
$$\vert N_{l_0+m, l_1, j}\,-\,  N_{\infty, l_1, j}\vert \, \leq \, 2,\quad \forall  \, j\in \{1, \dots, p(l_1)\}.$$
Thus for $m$  big enough,  the rotation number  $\rho_0$ of the configuration  ${(\theta_{l_0+m, n})}_n$  satisfies:
$$\frac{1}{\sum\limits_{j=1}^{p_{l_1}}\nu_{l_1,j}N_{l_0+m, l_1, j} + 2\nu(\Cc_{l_1})}\,\leq\, \rho_0\, \leq \, 
\frac{1}{\sum\limits_{j=1}^{p_{l_1}}\nu_{l_1,j}N_{l_0+m, l_1, j} }.$$
On the other hand 
$$\frac{1}{\sum\limits_{j=1}^{p_{l_1}}\nu_{l_1,j}N_{\infty, l_1, j} + 2\nu(\Cc_{l_1})}\,\leq\, \rho_\infty\, \leq \, 
\frac{1}{\sum\limits_{j=1}^{p_{l_1}}\nu_{l_1,j}N_{\infty, l_1, j} }.$$
This implies:
$$\left| \frac{1}{\rho_\infty}\, -\, \frac{1}{\rho_0}\right| \, \leq\, 8\nu(\Cc_{l_1}).$$
Since this last inequality is true for all $l_1 >l_0$, we get:
$$\rho_\infty\, =\, \rho_0.$$
\end{proof}

This ends the proof of Proposition \ref{18}.
\end{proof}
In order to prove  Part $(iii)$ of Theorem \ref{GGP}, we  choose a positive real number $\rho$ and consider a sequence 
of minimal configurations ${(\theta_{m, n})}_n$, $m\geq 0$,  with rotation number $\rho_m\in \Ff$  such that:
$$\lim_{m \to +\infty} \rho_m\, =\, \rho.$$
A discussion completely similar to the one we used in the proof of Lemma \ref{grec} allows us to show that there exists  
$M>0$ such that:
$$0\leq \theta_{m, n+1}\, -\, \theta_{m, n}\, \leq \, M\quad\quad  \forall \, m\geq 0, \quad \forall\, n\in \mz.$$
Consequently,  the set of all the   configurations  $(\theta_{m, n})_n$, for $m\geq 0$  and  their translated, is in a compact subset of $\mr^\mz$ and thus, as done previously, we can exhibit a subsequence of configurations which converges to a minimal configuration ${(\theta_n)}_n$. Thanks to  continuity property of the rotation number (Part $(ii)$ of Theorem \ref{GGP}), we conclude that the rotation number of ${(\theta_n)}_n$ is $\rho$.
\section{Final remarks}\label{fin}
\subsection {Dynamical systems}

Minimal configurations of the Frenkel-Kontorova   model  obviously satisfy the variational equations:
$$ U'(\theta_{n} - \theta_{n+1})\,-\, U'(\theta_{n-1} - \theta_{n}) \, + V'(\theta_n) \, =\, 0,\quad \forall n\in \mz.$$
By introducing the new variables\footnote{Recall that $U'$ is an increasing homeomorphism of the real line.}:
$$p_n\, =\, U'(\theta_{n-1} - \theta_{n}), \quad  \forall \, n\in \mz,$$ we get the dynamical system defined on $\mr\times \mr$ by:

$$
\left\{
\begin{array}{l}
p_{n+1}\, =\, p_n  \, -\, V'(\theta_n)\\
 \\
\theta_{n+1}\, =\, \theta_n\, - \, (U')^{-1}(p_n  \, -\, V'(\theta_n))\\
\end{array}
\right.
\qquad(\star\star\star)
$$

In the crystal case, $V'$ is a periodic function with period $L,$ the period of the crystal.
It follows that the map defined by $(\star\star\star)$ descends  to a  map on the open annulus 
$\mr/L.\mr\times\mr$ which is an orientation preserving  diffeomorphism which preserves the standard area form. 
Area preserving maps of the annulus have been widely studied and Aubry-Mather theory which makes a bridge 
between the Frenkel-Kontorova model and dynamical systems,  has been a powerful tool for both sides.

In the quasicrystal case, the dynamical system extends to  an area preserving "diffeomomorphism" \footnote{We mean diffeomomorphism in the leaf direction.}  on the solenoidal annulus $\Omega(\Ss)\times \mr.$  The study of such maps will be the subject of a forthcoming paper. 

\subsection {Quasicrystals in $\mr^d$, $\, d>1$} As we already noticed, the construction of the hull of a quasicrystal and its interpretation as an inverse limit of branched manifolds  can be done for quasicrystals in any dimension (see \cite{BG}, \cite{BBG}, \cite{Sadun}).  On the other hand, in a recent work \cite{KLR} ,  H. Koch, R de la Llave and C. Radin developed a generalization of Aubry-Mather theory for functions on lattices in $\mr^d$.  Both arguments make tempting to develop in a same way, a Aubry-Mather theory for quasicrystals in $\mr^d$, $\, d>1.$

\bigskip
\noindent  {\bf Acknowledgments:} {\it It is a pleasure for the authors to thank P. le Calvez for very helpful comments about  Aubry-Mather theory 	and an unknown referee for his useful  suggestions.  S. P.  has been supported  by ECOS-Conicyt grant C03-E03.}

\end{document}